\documentclass[twocolumn,showpacs]{revtex4}

\usepackage{graphicx}
\usepackage{dcolumn}
\usepackage{bm}

\begin{document}

\preprint{APS/123-QED}

\title{A molecular dynamics simulation of polymer crystallization \\
from oriented amorphous state}
\author{Akira Koyama$^1$, Takashi Yamamoto$^2$, Koji Fukao$^3$, and 
Yoshihisa Miyamoto$^4$}
\affiliation{$^1$Graduate School of Human and Environmental Studies,
Kyoto University, Kyoto 606-8501 Japan,\\ $^2$Faculty of Science,
Yamaguchi University, Yamaguchi 753-8512 Japan,\\ $^3$Department of
Polymer Science and Engineering, Kyoto Institute of Technology,
Matsugasaki, Kyoto 606-8585 Japan,\\ $^4$Faculty of Integrated Human
Studies, Kyoto University, Kyoto 606-8501 Japan}

\date{\today}
             
\begin{abstract}
Molecular process of crystallization from an oriented amorphous state
 was reproduced by molecular dynamics simulation for a realistic
 polyethylene model.  Initial oriented amorphous state was obtained by
 uniaxial drawing an isotropic glassy state at 100 K.  By the
 temperature jump from 100 K to 330 K, there occurred the
 crystallization into the fiber structure, during the process of which
 we observed the developments of various order parameters.  The real
 space image and its Fourier transform revealed that a hexagonally
 ordered domain was initially formed, and then highly ordered
 crystalline state with stacked lamellae developed after further
 adjustment of the relative heights of the chains along their
 axes.
\end{abstract}



\maketitle

The molecular dynamics (MD) simulation has recently come to be
recognized as a very promising tool to understand crystallization in
systems of chain molecules, since they can directly provide us with
molecular trajectories that are experimentally hard to access.  In an
early paper Rigby and Roe successfully reproduced a highly ordered
lamella structure from a bulk amorphous state of {\it n}-decane
\cite{RIGBYandROE}.  The work was followed by studies that elaborated on
detailed crystallization process in bulk {\it n}-alkanes of various
length \cite{ESSELINK,TAKEUCHI}.  Crystallization in thin films of {\it
n}-alkanes was also investigated using MD technique
\cite{SHIMIZUandYAMAMOTO,LIandYAMAMOTO}.   Recent advances have come to
enable the quantitative estimation of the growth rate 
vs. crystallization temperature~\cite{WAHEEDandLAVINEandRUTLEDGE}.

Chain folded crystallization of polymers has also been investigated by
molecular simulations.  Kavassalis and Sundararajan first reported the
MD studies on the collapse and folding of a single polyethylene molecule
in vacuum \cite{KAVASSALISandSUNDARARAJAN}, while Muthukumar and
coworkers investigated a primary nucleation and subsequent crystal
growth in solution by Langevin dynamics simulation
\cite{LIUandMUTHUKUMAR,WELCHandMUTHUKUMAR}. The molecular process of
secondary nucleation and growth of lamella crystals has also been a
subject of great interest and several investigations have been reported
\cite{LIUandMUTHUKUMAR,YAMAMOTO1,DOYEandFRENKEL}.  We can notice in very
recent literatures a surge of interest in reproducing chain folded
crystallization from the melt and solution both during the primary
nucleation \cite{WELCHandMUTHUKUMAR,MEYERandMULLER-PLATHE} and the
growth \cite{YAMAMOTO2}, though at present we can only treat relatively 
short polymer chains of about hundreds of segments.

Crystallization of much longer chains in the bulk system is expected to
require extraordinary long simulation time even by use of present day
supercomputers, especially when we adopt realistic molecular models.  We
have already reported an MD simulation of polymer ordering in the bulk
polyethylene \cite{KOYAMA}.  We could there observe only the local
ordering in chain packing, and we discussed the results within the
context of recent experimental and theoretical investigations on
precursor phenomena prior to 
crystallization~\cite{IMAIandKAJI,FUKAOandMIYAMOTO,SASAKIandTASHIRO,OLMSTED}.

The polymeric material needs very long time until the onset of
crystallization and can be easily quenched into amorphous glassy state
with isotropic chain orientation.  By drawing the amorphous sample below
the glass transition temperature, we obtain the oriented amorphous state
in which the chains run parallel to the draw direction on the average.
We can then expect to accelerate the crystallization if we use such
oriented amorphous sample, instead of isotropic melt \cite{KOYAMA}, as
an initial state of MD simulation.  In fact, recent experiments revealed
that the crystallization rates of polymers depend strongly on the degree
of the chain orientation \cite{MAHENDRASINGAM}.  Moreover, it will be
found that the oriented amorphous state is very convenient in
investigating the spontaneous development of fiber structures with
stacked lamellae or the detailed molecular level structure of the
amorphous phase between 
lamellae~\cite{GAUTAMandBALJEPALLIandRUTLEDGE}.

In this Letter we report results of our recent simulation of polymer
crystallization from bulk oriented amorphous state.  The molecular model
we adopted was essentially the same as that of our previous work
\cite{KOYAMA}.  We considered the energies of bond stretching, bond
angle bending, and torsion of the united atom model of polyethylene
(PE).  The truncated Lennard-Jones 12-6 potential was used to reproduce
van der Waals interactions between the atoms, where a rather short
cutoff radius of 0.85 nm was used to improve computational efficiency.
Bulk polymer system was made with a single linear PE composed of 5000
united atoms being placed in the MD cell under periodic boundary
condition.  Newton's equations of motion were integrated by the
leap-flog algorithm with a time step of 4 fs, and the temperature was
controlled every 10 steps by the ad-hoc velocity rescaling, while the
pressure was adjusted every 100 steps by the loose-coupling 
method~\cite{BERENDSEN}.  The parameter M used in the pressure control was 
set to be 13.26 Pa$\cdot$s/m, and the shape of the MD cell was kept 
orthorhombic.  

\begin{figure}
\includegraphics[width=8cm,keepaspectratio]{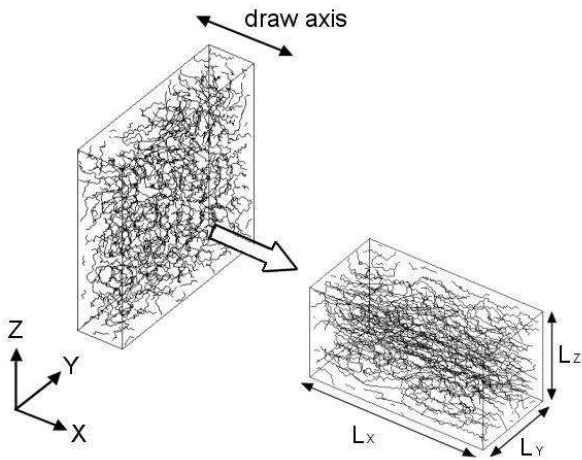}
\caption{Schematic picture of the preparation of oriented amorphous sample.}
\label{fig1}
\end{figure}

   First we prepared the isotropic melt as follows.  An initial
configuration of the chain, with fixed bond lengths and bond angles, was
generated using the technique developed by Theodorou and Suter
\cite{THEODOROUandSUTER}, and then a short MD simulation of 10 ps under
constant volume and temperature (600 K) was made to reduce the overlap
of atoms. This initial state was then fully relaxed for 1 ns at constant
temperature (600 K) and pressure (1 atm).

Next we generated the oriented amorphous state by drawing the isotropic
amorphous sample well below the glass transition temperature. The glassy
amorphous sample was drawn at 100K along the x-axis at a constant rate
of 500 \% /ns up to a final draw ratio of 400 \% (Fig.1), where the
method of uniaxial drawing was the same as that used by Ogura and
Yamamoto \cite{OGURAandYAMAMOTO}.  During drawing, the temperature was
kept constant at 100 K and the lateral widths of the cell were allowed
to shrink along the directions perpendicular to the draw axis under
constant pressure (1 atm).  After attaining the final draw ratio of
400\%, further relaxation was carried out for 1.5 ns 
under constant pressure (1 atm)  and temperature (100 K).

Crystallization of the oriented amorphous state was then simulated for
30 ns, under constant temperature (330 K) and pressure (1 atm).  Figure
2 shows the changes with time of various quantities of the system.  The
internal energy slowly decreases until 10 ns (Fig. 2 (a)), but shows the
steep reduction around 12 ns suggesting an onset of ordering.  The
specific volume also decreases almost lineally up to around 14 ns
(Fig. 2 (b)), where the MD-cell shows clear initial shrink within 4 ns
and then expands along the draw axis (Fig. 2 (c)) at the most pronounced
rate around 12 ns.  Around 15 ns, the MD-cell dimensions nearly come to
saturate.  It might seem strange that the specific volume increases
again after 15 ns, but later this will be found an artifact due to the
imposed periodic boundary condition and very slow relaxation in the well
developed ordered state.

PE molecule has the lowest energy conformation trans and the second
lowest gauche$^\pm$.  Since the molecule has all trans conformation in
the crystal, we evaluated the chain ordering by use of the trans
fraction and the mean length of the trans sequence; the trans and the
gauche bonds are here defined as the bonds whose torsion angles
distribute around 0$^\circ$ and $\pm$120$^\circ$, respectively, within
the range of 60$^\circ$, whereas the length of the trans sequence is
expressed by the number of the trans bonds in the sequence.  The
transition rates in Fig. 2(d) slowly decrease until around 10 ns, but
then they abruptly fall around 12 ns until the transitions nearly halt
around 17 ns.  Both the trans fraction and the mean length of the trans
sequences in Figs. 2 (e) and 2 (f) increase steadily and show a marked
jump up around 12 ns until they saturate completely around 17 ns.  The
final state obtained after about 20 ns is thus considered to be made
predominantly of the trans segments and highly ordered with very
infrequent internal rotations between the trans and gauche.

In order to investigate the development of orientational order of the
molecule, we examined the orientation of the C-C-C segments and the
correlation volume that is defined as a measure of parallel alignment of
the segments \cite{RIGBYandROE}.  The orientations of the C-C-C segments
along the three axes are evaluated by the mean squared cosines
$<\cos^2\theta_X>$, $<\cos^2\theta_Y>$, and $<\cos^2\theta_Z>$, where
${\theta s}$   are the angles between the orientation vector spanning
every other carbons of the C-C-C segment and the axes.  The correlation 
volume is defined as
\begin{eqnarray}
<V_{c}> = < \int^{r_{\rm max}}_{0} 4 \pi r^{2} g(r)  s(r) dr >,
 \nonumber
\end{eqnarray}
where $g(r)$ is the radial distribution function of the center of mass
for the segment, and $s(r)$ is the orientation correlation function
between two segments defined as $s(r)=1/2(3<\cos^2
\theta_{i,j}(r)>_{i,j})$, where $\theta_{i,j}(r)$  is the angle between
the $i$-th and the $j$-th orientation vectors separated by $r$, and the
average is taken over all pairs within the shell $r \sim r+\Delta r$,
and the upper limit of the integration $r_{max} =$ 1.6 nm.  The chain
orientation along the draw axis and the correlation volume were
initially unstable, but started to increase around 4 ns until they
saturate around 20 ns (Figs. 2(g), 2(h)).  The rapid increase in the
correlation volume around 12 ns indicates the spontaneous orientation
and the parallel alignment of the segments.

\begin{figure*}
\includegraphics[width=15cm,keepaspectratio]{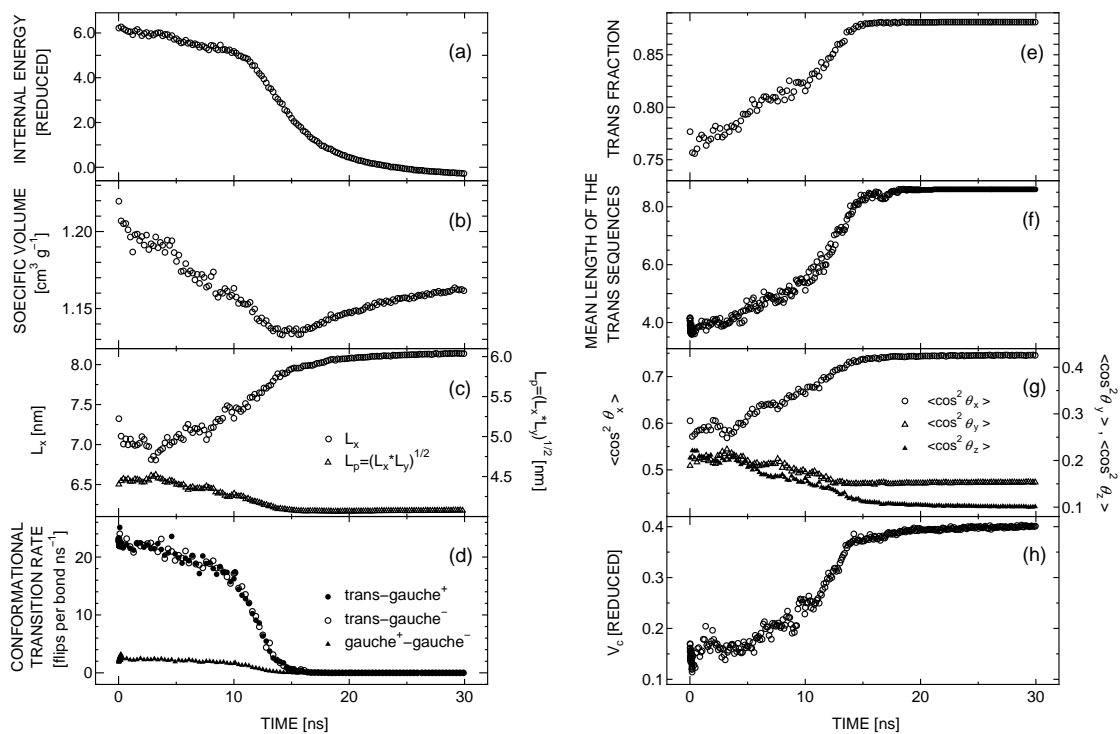}
\caption{Changes in the various quantities with time: (a) the internal
 energy, (b) the specific volume, (c) the side length of the MD cell,
 (d) the conformational transition rate, (e) the trans fraction, (f) the
 mean length of the trans sequences, (g) the degree of the orientation,
 (h) the correlation volume.}
\label{fig2}
\end{figure*}

\begin{figure*}
\includegraphics[width=17cm,keepaspectratio]{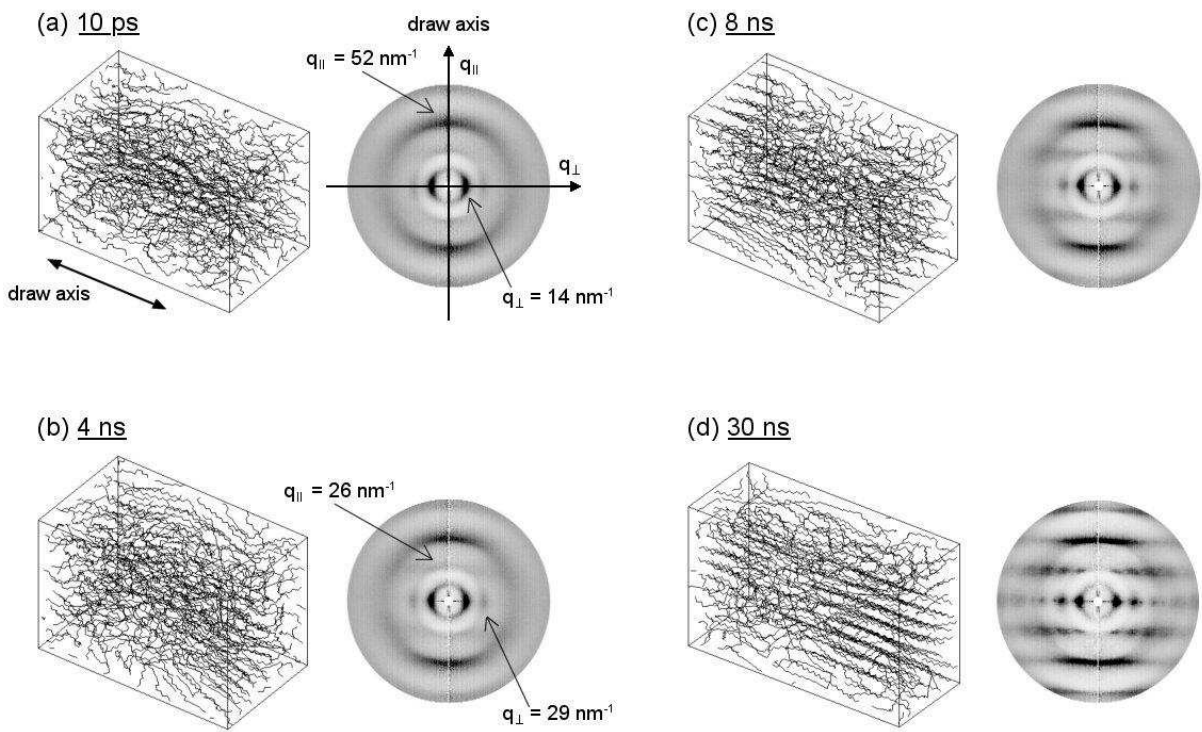}
\vspace*{-0.5cm}
\caption{Snapshots (left), and the structure functions (right) 
averaged around the draw axis, $S_{\rm 2D}(q_{||},q_{\perp})$. }
\label{fig3}
\end{figure*}

The structural changes described above were analyzed by use of
real-space image and its Fourier transform, the structure function.
Three dimensional structure function is defined by
\begin{eqnarray}
S(\bf{q}) &=& \frac{1}{N} \int\!\int d{\bf r}d{\bf r'} <\rho ({\bf r}) \rho({\bf r'})> \exp (-i {\bf q} \cdot ({\bf r}-{\bf r'})),
 \nonumber
\end{eqnarray}
where ${\bf q}$ is the wave vector, ${\bf r}$ and ${\bf r'}$ are the 
position vectors, $N$ is the total number of atoms. 
The atomic density at the position ${\bf r}$, $\rho({\bf r})$, is
defined by the relation $\rho({\bf r}) = \sum_{j=1}^N \delta ({\bf
r}-{\bf r}_{j})$, where ${\bf r}_{j}$ represents the position of the
$j$-th atom.  Here we consider the structure function averaged around
the draw axis, $S_{\rm 2D}(q_{||}, q_{\perp})$, where $q_{||}$ and
$q_{\perp}$ are the components of the wave vector along the draw axis
and perpendicular to it, respectively.  The intensity profile at 10 ps
(Fig. 3 (a)) indicates that the system is initially in the oriented
amorphous state.  We readily notice reflections on the equator at
$q_{\perp}=14~{\rm nm}^{-1}$ which are attributed to the nearest
neighbor interchain length, and those on the meridian at $q_{||}=52~{\rm
nm}^{-1}$ which correspond to half the fiber period of the polymethylene
chain.  At 4 ns, Fig. 3(b), broad layer streaks are visible along the
layer line at $q_{||}=26~{\rm nm}^{-1}$, and the reflections at
$q_{\perp}=29~{\rm nm}^{-1}$ on the equator are slightly sharper and
stronger than in Fig. 3(a).  It is clearly found in the snapshots that
the degree of the order at 4 ns is considerably higher than that at 10
ps.  At 8 ns, Fig. 3 (c), the layer streaks at $q_{||}=26~{\rm nm}^{-1}$
become much more evident, and the equatorial reflections around
$q_{\perp}=29~{\rm nm}^{-1}$ come to sprit.  Such diffraction pattern is
quite similar to that observed in the hexagonal phase of PE, which
appears under high pressure and temperature \cite{YAMAMOTO3} or under
high extension \cite{PENNINGSandZWIJNENBURG}.  The snapshot in Fig. 3
(c) shows that the nearby chains tend to orient along the draw axis and
form crystalline domains, which emerged around 4 ns and gradually became
lager.  After 10 ns, marked intensity contrast appears on the streak at
$q_{||}=26~ {\rm nm}^{-1}$, and finally at 30 ns the diffraction come to
show ordered crystalline pattern (Fig. 3 (d)) showing the development of
high longitudinal order in the crystal.

Around 15 ns the well-developed crystalline layer nearly filled the
system.  With further development of order, the amount of amorphous
segments decreases.  But, due to the periodic boundary condition,
several parts of the layer surface contacted with the opposite layer
surface of itself, which prevent the shrinkage of the MD cell along the
draw direction and gives rise to the density decrease in the amorphous
phase (Figs. 2 (b), 3 (d)). 

In summary, we have performed the MD simulations of the polymer
crystallization from the oriented amorphous states.  We have succeeded,
for the first time, in reproducing the fiber structure of stacked
lamellae with ordered crystalline state.  We have observed that, after
initial relaxation for about 4 ns, the hexagonal order first appears and
grows until around 10 ns, and then develops the highly crystalline state
through longitudinal adjustments of the parallel chains in the crystal.
Analysis of the fiber structure obtained, such as the conformation of
tie chains and the structure of the interlamellar amorphous phase, and
the detailed molecular process of fiber formation are also very
important problems, and they will be published in a separated paper.

This work was supported by a Grant-in-Aid for Science Research on
Priority Areas, "Mechanism of Polymer Crystallization" (Grants
Nos. 12127203, 12127204 and 12127206) from The Ministry of Education, 
Science, Sports and Culture of Japan.


\end{document}